\documentclass[12pt]{article}
\usepackage{amsmath,amssymb,epsfig}
\usepackage{overpic}
\usepackage{graphicx,floatflt,subfigure}

\usepackage{color}
\input{colordvi.tex}

\renewcommand{\thefootnote}{\fnsymbol{footnote}}

\newcommand{\tr}{{\rm Tr}}
\newcommand{\del}{\partial}
\newcommand{\fphi}{\langle F_\phi \rangle}
\newcommand{\vs}{\langle S \rangle}
\newcommand{\fs}{\langle F_S \rangle}
\newcommand{\vx}{\langle X \rangle}

\makeatletter

\renewcommand\section{\@startsection {section}{1}{\z@}%
                                   {-3.5ex \@plus -1ex \@minus -.2ex}%
                                   {2.3ex \@plus.2ex}%
                                   {\normalfont\large\bfseries}}

\renewcommand\subsection{\@startsection{subsection}{2}{\z@}%
                                     {-3.25ex\@plus -1ex \@minus -.2ex}%
                                     {1.5ex \@plus .2ex}%
                                     {\normalfont\normalsize\bfseries}}

\newcount\hour \newcount\minute
\hour=\time \divide \hour by 60
\minute=\time
\def\now{%
\ifnum \hour<13
  \ifnum \hour=0 \advance \hour by 12 \number\hour:\else \number\hour:\fi%
     \ifnum \minute<10 0\fi%
     \number\minute%
\ A.M.%
\else \advance \hour by -12 \number\hour:%
  \ifnum \minute<10 0\fi%
  \number\minute%
  \ P.M.%
\fi%
}

\makeatother

\begin{document}

\baselineskip=18pt  
\numberwithin{equation}{section}  
\allowdisplaybreaks  



%
%


\thispagestyle{empty}

\vspace*{-2cm}
\begin{flushright}
\end{flushright}

\begin{flushright}
KUNS-2326 \\
YITP-11-37 \\
\end{flushright}

\begin{center}

\vspace{2.0cm}

{\bf\Large (Extra)Ordinary Gauge/Anomaly Mediation}
\\

\vspace*{1.5cm}
{\bf
Tatsuo Kobayashi$^{1}$\footnote{e-mail: {\tt kobayash@gauge.scphys.kyoto-u.ac.jp}}, Yuichiro Nakai$^{2}$\footnote{e-mail:
{\tt ynakai@yukawa.kyoto-u.ac.jp}} and Manabu Sakai$^{2}$}\footnote{e-mail: {\tt msakai@yukawa.kyoto-u.ac.jp}} \\
\vspace*{0.5cm}

$^{1}${\it {Department of Physics, Kyoto University, Kyoto 606-8502, Japan}} \\
\vspace{0.1cm}

$^{2}${\it {Yukawa Institute for Theoretical Physics, Kyoto University, Kyoto 606-8502, Japan}}
\vspace{0.1cm}

\end{center}

\vspace{1cm} \centerline{\bf Abstract} \vspace*{0.5cm}

We study anomaly mediation models with gauge mediation effects from
messengers which have a general renormalizable mass matrix with a
supersymmetry-breaking spurion. 
Our models lead to a rich structure of supersymmetry breaking terms 
in the visible sector.
We derive sum rules among the soft scalar masses for each generation.
Our sum rules for the first and second generations are the same 
as those in general gauge mediation, but 
the sum rule for the third generation is different because of 
the top Yukawa coupling.
We find the parameter space
where the tachyonic slepton problem is solved. We also explore the
case in which gauge mediation causes the anomalously small gaugino
masses. Since anomaly mediation effects on the gaugino masses exist,
we can obtain viable mass spectrum of the visible sector fields.

\newpage
\setcounter{page}{1} 
\setcounter{footnote}{0}
\renewcommand{\thefootnote}{\arabic{footnote}}





\section{Introduction}

Supersymmetry (SUSY) is one of the promising candidates for physics beyond the standard model. It can solve the hierarchy problem on the Higgs mass. In addition, the standard model gauge couplings are unified at a high-energy scale in the minimal supersymmetric extension of the standard model (MSSM). Moreover, the lightest supersymmetric particle (LSP) is a strong candidate for the dark matter in our universe.

If SUSY is realized in nature, it must be broken at an energy scale
above the weak scale. We usually leave its dynamics to the hidden
sector different from our visible sector. Anomaly-mediated
supersymmetry breaking \cite{Randall:1998uk, Giudice:1998xp} is
one of the attractive mechanisms to transmit the SUSY breaking of the
hidden sector into the visible sector fields without flavor
problems. In anomaly mediation, the SUSY breaking in the hidden
sector is encoded in the F-component of the superconformal compensator
$\phi = 1 + \theta^2 \fphi$. The soft SUSY breaking parameters in the
visible sector are given by the 1-loop suppressed form of the
parameter $\fphi$. The gravitino naturally has  $\mathcal{O}(10) \,
\text{TeV}$ of a mass. 
Unfortunately, anomaly mediation suffers from the problem
that the slepton masses become tachyonic. Then, we need to modify the
original form by some additional effects 
in order to obtain a successful model \cite{Pomarol:1999ie,Chacko:1999am}.

On the other hand, in gauge-mediated supersymmetry breaking
\cite{GaugeMediation, Giudice:1998bp}, the SUSY breaking
of the hidden sector is transmitted into the visible sector by the
standard model gauge interactions. As in anomaly mediation, the
unwanted flavor-changing processes are suppressed due to the flavor
blindness of the gauge interactions and hence gauge mediation is
also considered to be a promising mediation mechanism of the SUSY
breaking. However, in gauge mediation, we often encounter the
anomalously small gaugino masses compared to the scalar masses. In
this case, we cannot obtain ${\cal O}(1) \, \text{TeV}$  of the gaugino masses and the scalar masses at the same time.\footnote{The anomalously small gaugino mass problem can be seen in direct gauge mediation models \cite{Izawa:1997gs,Komargodski:2009jf} (see also \cite{Nakai:2010th}) and semi-direct gauge mediation models \cite{Semidirect}.}
That may cause a hierarchy problem, again.

Both anomaly mediation and gauge mediation are quite
interesting, but both may have problems.
In particular, the pure anomaly mediation has the problem 
in the slepton sector, while gauge mediation may have 
the problem in the gaugino masses.
Then, a natural approach to these problems would be to mix anomaly
mediation and gauge mediation
\cite{Pomarol:1999ie,Nelson:2002sa,Hsieh:2006ig,Cai:2010tj,Sundrum:2004un}.
The contributions to the soft masses from anomaly mediation and gauge mediation can be naturally comparable when we derive the mass scale of the messengers of gauge mediation from the parameter $\fphi$. The tachyonic slepton problem in the pure anomaly mediation can be cured by the contribution from gauge mediation while the anomalously small gaugino masses can be enhanced by the anomaly mediation contribution.

In this paper, we extend the model of \cite{Hsieh:2006ig}, which mix anomaly mediation and gauge mediation, to models with the following generalized fermion mass matrix of the messenger fields \cite{Cheung:2007es}:
\begin{equation}\label{model}
\mathcal{L}_{mess} = \int d^2 \theta \, \phi \mathcal{M}_{ij} (X) \psi_i \tilde{\psi}_j + h.c. = \int d^2 \theta \, \phi ( \lambda_{ij} X + m_{ij}) \psi_i \tilde{\psi}_j + h.c.,
\end{equation}
where $X = \langle X \rangle + \theta^2 \langle F_X \rangle$ is a SUSY
breaking spurion and $\lambda_{ij}$, $m_{ij}$ are constant matrices,
which are in general independent of each other. 
The fields $\psi_i, \tilde{\psi}_j \,\, (i = 1, \cdots, N)$ denote the messengers which belong to the (anti-)fundamental representations under the $SU(5)$ group into which the standard model gauge symmetry is embedded.
We can further generalize the messenger sector such that the superpotential has the doublet/triplet splitting as follows:
\begin{equation}
\begin{split}
\mathcal{L}'_{mess}  
&= \int d^2 \theta \, \phi \left[\mathcal{M}_{ij}^2 (X) \ell_i \tilde{\ell}_j + \mathcal{M}_{ij}^3 (X) q_i \tilde{q}_j \right] + h.c. \\
&= \int d^2 \theta \, \phi \left[( \lambda_{2ij} X + m_{2ij}) \ell_i \tilde{\ell}_j + ( \lambda_{3ij} X + m_{3ij}) q_i \tilde{q}_j \right] + h.c.,
\end{split}\label{model2}
\end{equation}
where $\ell_i, \tilde{\ell}_j$ and $q_i, \tilde{q}_j$ are $SU(2)$ doublets and $SU(3)$ triplets of the messengers respectively.
We here take this general case with some conditions for simplicity.
In this model, we study the soft mass spectrum of the model and identify the LSP. 

The rest of the paper is organized as follows. In section 2, we will
present our model which gives the messenger mass matrices
\eqref{model}, \eqref{model2} and analyze the soft mass spectrum of
the visible sector fields. In section 3, we will show the numerical
analyses of the soft masses. In addition, we will derive sum rules among the soft scalar masses for each generation.
In section 4, we will conclude the discussions.

\section{Generalities}

In this section, we first show a model which leads to the messenger mass matrix \eqref{model}, \eqref{model2} and analyze its vacuum structure,  following the discussion of \cite{Hsieh:2006ig}. Then, we present the soft mass formulae of the visible sector fields derived from anomaly mediation and gauge mediation.

\subsection{The models}

In addition to the usual canonical K\"ahler potential, 
we consider the following terms in Lagrangian of the messenger fields $\psi_i, \tilde{\psi}_j$ and the singlet field $S$:
\begin{equation}\label{lag}
\Delta \mathcal{L} = \int d^4 \theta \, \frac{\phi^\dagger}{\phi} \left( \frac{1}{2} c_S S^2 + c_{P ij} \psi_i \tilde{\psi}_j \right) + \int d^2 \theta \left[ \frac{\lambda_S}{3!} S^3 + \lambda_{P ij} S \psi_i \tilde{\psi}_j \right] + h.c.,
\end{equation}
where $c_S, \lambda_S, c_{P ij}$ and $\lambda_{P ij}$ are the real
coupling constants. 
Here, we assume that the quadratic terms of $\psi_i, \tilde{\psi}_j$ and
$S$ are absent.\footnote{We can forbid these terms by a discrete $R$
  symmetry such as $S (\theta ) \mapsto - S (i\theta ), \psi_i (\theta )
  \mapsto - \psi_i (i\theta ), \tilde{\psi}_j (\theta ) \mapsto -
  \tilde{\psi}_j (i\theta )$ with the other fields even \cite{Hsieh:2006ig}.}
{}From the above Lagrangian, the scalar potential of this model is given by
\begin{equation}
\begin{split}
V &= \left| c_S \langle F_\phi^\dagger \rangle S + \frac{1}{2} \lambda_S S^2 + \lambda_{P ij} \psi_i \tilde{\psi}_j \right|^2 \\
&\quad+ \left| \left( c_{P ij} \langle F_\phi^\dagger \rangle + \lambda_{P ij} S \right) \psi_i \right|^2 + \left| \left( c_{P ij} \langle F_\phi^\dagger \rangle + \lambda_{P ij} S \right) \tilde{\psi}_j \right|^2 \\
&\quad+ \left| \langle F_\phi \rangle \right|^2 \left( \frac{1}{2}c_S S^2 + c_{P ij} \psi_i \tilde{\psi}_j \right)+ h.c.
\end{split}\label{potential}
\end{equation}
We next consider the minimization of this potential. We assume that
the messenger fields $\psi_i, \tilde{\psi}_j$ stabilize at the origin of
their field space, $\langle \psi_i \rangle = \langle \tilde{\psi}_j
\rangle = 0$ to preserve the standard model gauge symmetry. The
expectation values of $S$ and $F_S$ on the minimum of the potential
are then given by\footnote{See \cite{Hsieh:2006ig, Cai:2010tj} for a
  detail of the potential analysis. In \cite{Cai:2010tj}, it is pointed
  out that there is a UV divergent 1-loop linear term of S. 
However, such a term does not affect our results, because we parametrize 
$\langle S \rangle$ and $\langle F_S \rangle$ for our phenomenological
purpose.}
\begin{equation}\label{vev}
\begin{split}
\langle S \rangle = - \frac{\fphi}{2\lambda_S} \left( 3c_S + \sqrt{c_S(c_S-8)} \right), \\
\Big\langle \frac{F_S}{S} \Big\rangle = \frac{\fphi}{4} \left( -c_S + \sqrt{c_S(c_S-8)} \right).
\end{split}
\end{equation}
Note that this vacuum is the global minimum of the potential in a certain parameter range.
Then, these vacuum expectation values lead to the following mass term for the messenger fields $\psi_i, \tilde{\psi}_j$:
\begin{equation}\label{mass}
\mathcal{L}'_{mess} = \int d^2 \theta \, \phi \left( M_{ij} + F_{ij} \theta^2 \right) \psi_i \tilde{\psi}_j,
\end{equation}
where we define
\begin{equation}
\begin{split}
M_{ij} &= \fphi c_{P ij} + \vs \lambda_{P ij}, \\
F_{ij} &= -2\fphi M_{ij} + \left(\fphi \vs + \fs \right) \lambda_{P ij}.
\end{split}
\end{equation}
Note that the matrix $M_{ij}$ is not proportional to the matrix
$F_{ij}$ in general. Thus, we can obtain the form of the mass term of
the messengers \eqref{model}. The messenger scale is naturally the
same order as the scale $\fphi$ when all the couplings in the model
are of $\mathcal{O}(1)$ and hence the anomaly-mediated and the
gauge-mediated contributions are comparable. If we tune the
parameters, we can realize the cases where the effect of anomaly
mediation is dominant or that of gauge mediation is dominant. 
Since the vacuum
we consider here is the global minimum of the potential
\eqref{potential}, we may worry about the consistency with the
discussion of \cite{Komargodski:2009jf} where the pseudomoduli space
of the SUSY breaking vacuum cannot be locally stable everywhere in
order to generate sizable gaugino masses in direct gauge
mediation. However, there is no pseudomoduli space in our set-up and
hence the discussion of \cite{Komargodski:2009jf} cannot be applied as
discussed in \cite{Nakai:2010th}. In this case, we can obtain sizable
gaugino masses in the global minimum of the potential. The models
discussed in \cite{Nakai:2010th,Nomura:1997ur,Ibe:2010jb} with the minimal gauge mediation \cite{Dine:1994vc} have the messenger sector separated from the SUSY breaking sector and they have the additional messenger gauge interaction (or the nonrenormalizable interaction in the K\"ahler potential) between these two sectors. Then, these models do not have a pseudomoduli space in the messenger sector and realize sizable gaugino masses in the global vacuum. In the present set-up, the messenger gauge interaction in the models of \cite{Nakai:2010th,Nomura:1997ur} is replaced to the interactions of the conformal compensator field $\phi$ with the messenger sector fields. Then, the SUSY breaking in the messenger sector mediated by the compensator can generate nonzero leading order gaugino masses in the global minimum of the potential.

We here comment on the difference between the model in \cite{Cheung:2007es} and our set-up. In the model of \cite{Cheung:2007es}, the nontrivial R-charge assignment on the SUSY breaking field $X$ and the messenger fields $\psi_i, \tilde{\psi}_j$ restricts the determinant of the matrix $\mathcal{M} (X)$ to the following form:
\begin{equation}
\det \mathcal{M} = X^n G(m, \lambda), \quad n = \frac{1}{R(X)} \sum^N_{i=1} (2 - R(\psi_i) - R(\tilde{\psi}_i)),
\end{equation}
where $G(m, \lambda)$ is some function of the coupling constants $m,
\lambda$ and $R(X), R(\psi_i), R(\tilde{\psi}_j)$ are the R-charges of the
fields $X, \psi_i, \tilde{\psi}_j$. Then, when we introduce the
doublet/triplet splitting into the messenger sector but take the same
R-charge assignments of the $SU(2)$ doublet and $SU(3)$ triplet parts
of the messengers, the following  GUT relation among the gaugino masses is preserved:
\begin{equation}
M_1 : M_2 : M_3 = \alpha_1 : \alpha_2 : \alpha_3,
\end{equation}
where $M_1, M_2, M_3$ are the masses of the bino, wino and gluino fields respectively.
On the other hand, in our model, any condition is not imposed on the
coupling constants $c_S, \lambda_S, {c_{P}}_{ij}$ and ${\lambda_{P}}_{ij}$. Then, we
have the general messenger mass matrix $\mathcal{M} (X)$ and the
determinant. Thus, in general, we do not have the above relation of
the gaugino masses when we introduce the doublet/triplet
splitting in the messenger sector. 
We can realize various ratios between the gaugino masses in our model.
For simplicity, in order to analyze numerically spectra of our models in 
section 3, 
we require the same structures for the mass matrices of the
$SU(2)$ doublet and $SU(3)$ triplet messengers $\ell_i, \tilde{\ell}_j$
and $q_i, \tilde{q}_j$ and preserve the GUT relation among the gaugino masses.

\subsection{The soft mass formulae}

We next show the formulae of the soft mass parameters of the visible sector derived from the anomaly mediation effects and the gauge mediation effects of the messengers whose fermion mass matrix is given by \eqref{model}.
First, we calculate the gaugino masses which can be read off from the holomorphic gauge coupling $\tau$ dependent on the spurion chiral superfield $X$ and the compensator chiral superfield $\phi$. We can write the gaugino mass as the following form:
\begin{equation}
M_\lambda = \frac{i}{2\tau} \left( \frac{\del \tau}{ \del \phi} \Big|_{\phi = 1} F_\phi + \frac{\del \tau}{\del X} \Big|_{X = \vx} F_X \right).
\label{gaugino}
\end{equation}
The holomorphic gauge coupling at a scale $\mu$ below the messenger scale is given by
\begin{equation}
\tau(\mu) = \tau_0 + i \frac{b'}{2\pi} \log{\frac{1}{\Lambda}} - \frac{i}{2\pi} \log{\text{det}\mathcal{M}} + i \frac{b}{2\pi} \log{\frac{\mu}{\phi}},
\end{equation}
where $b'$ is the $\beta$ function coefficient of the theory including the messenger fields while $b$ is the $\beta$ function coefficient in the effective theory below the mass scale of the messengers. The constant $\Lambda$ is the cutoff scale of the model and $\tau_0$ is the value of the coupling at that scale. 
Inserting this expression of the holomorphic coupling into \eqref{gaugino}, the gaugino mass is given by
\begin{equation}\label{gauginomass}
M_\lambda = \frac{\alpha}{4\pi} \left( b F_\phi + \frac{\del}{\del X} \log{\text{det}\mathcal{M}} \Big|_{X = \langle X \rangle} F_X \right),
\end{equation}
where $\alpha \equiv g^2/4\pi$. The first term is considered to be the anomaly-mediated contribution and the second term is the gauge mediation contribution.

We next derive the soft scalar mass of the matter field in the visible sector, which can be read off from the wavefunction renormalization factor,
\begin{equation}
Z = Z \left(\frac{\mu}{\Lambda|\phi|}, \frac{|X|}{\Lambda} \right),
\label{wave}
\end{equation}
which is the function of the combinations of $\mu/\Lambda|\phi|$ and $|X|/\Lambda$.
Then, the soft scalar mass of the matter field can be expressed as follows:
\begin{equation}\label{scalarmass2}
\begin{split}
m_Q^2 &= -\frac{1}{4} \frac{\del^2 \log{Z}}{\del (\log{\mu})^2} \left| F_\phi \right|^2 - \frac{1}{4} \frac{\del^2 \log{Z}}{\del (\log{|X|})^2} \left| \frac{F_X}{X} \right|^2 \\
&\hspace{1cm} + \frac{1}{4} \frac{\del^2 \log{Z}}{\del \log{\mu} \del \log{|X|} } F_\phi \frac{F_X^\dagger}{X^\dagger} + h.c. 
\end{split}
\end{equation}
The first term is considered to be the anomaly-mediated contribution and here we can replace the derivative of the compensator field $\phi$ to the derivative of the scale $\mu$ because of the factor dependence of the wavefunction renomalization factor \eqref{wave}. The second term is the gauge-mediated contribution. The rest terms are the mixing terms of both contributions and hence the final result of the soft scalar mass is not the simple sum of the anomaly-mediated effect and the gauge-mediated effect.
We deal with three contributions to the soft scalar mass one by one.
The first term of the anomaly-mediated contribution can be rewritten in terms of the anomalous dimension $\gamma$ and the $\beta$ function at the scale $\mu$ such as
\begin{equation}
-\frac{1}{4} \frac{\del^2 \log{Z}}{\del (\log{\mu})^2} \left| F_\phi \right|^2 = - \frac{1}{4}  \left( \frac{\del \gamma}{\del g} \beta_g + \frac{\del \gamma}{\del y} \beta_y \right) \left| F_\phi \right|^2,
\end{equation}
where $y$ is the Yukawa coupling.
At the 1-loop order, the anomalous dimension and the $\beta$ functions of the gauge coupling and the Yukawa coupling are given by
\begin{equation}
\begin{split}
\gamma &= \frac{1}{16\pi^2} \left( 4C_2g^2 - a y^2 \right), \\
\beta_g &= - \frac{bg^3}{16\pi^2}, \\
\beta_y &= \frac{y}{16\pi^2} \left( ey^2 - fg^2 \right),
\end{split}
\end{equation}
where $C_2$ is a quadratic Casimir and other coefficients in the above expressions are summarized in the Appendix. 
Next, we consider the rest of the terms in \eqref{scalarmass2}.
Here, we can ignore the $X$ dependence of the Yukawa coupling in the above expression of the anomalous dimension $\gamma$. We assume that the mass eigenvalues of the messenger fields are the same order and take a common messenger scale $M_{mess}$.
Then, we finally obtain the following soft scalar mass at the messenger scale $M_{mess}$:
\begin{equation}\label{scalarmass}
\begin{split}
m_Q^2 &= \left[ 2 b C_2 \left( \frac{\alpha}{4\pi} \right)^2 + \frac{1}{2} a \frac{y^2}{(4\pi)^2} \left( e \frac{y^2}{(4\pi)^2} - f\frac{\alpha}{4\pi} \right) \right] |F_\phi|^2 \\
&\quad \hspace{1cm} + 2C_2 \left( \frac{\alpha}{4\pi} \right)^2 \sum_i \left( \frac{\del \log{|a_i|}}{\del \log{|X|}} \right)^2 \left| \frac{F_X}{X} \right|^2 \\
&\quad \hspace{2cm} + 2 C_2 \left( \frac{\alpha}{4\pi} \right)^2 \frac{\del}{\del \log{|X|}} \log{|\text{det}\mathcal{M}|} F_\phi \frac{F_X^\dagger}{X^\dagger} + h.c.,
\end{split}
\end{equation}
where $a_i$ is the eigenvalue of the messenger mass matrix $\mathcal{M}$.
In appendix, we write down the explicit soft masses of the MSSM fields.
In the next section and Appendix, we set only the top Yukawa 
coupling $y_t$ non-vanishing, but the other Yukawa couplings vanishing.

\section{Spectrum and phenomenology}

In this section, we show the soft mass spectrum in the MSSM numerically by using the formula shown in the previous section.
We investigate two cases where the leading contribution to the gaugino mass from gauge mediation is zero or nonzero.
We also give a comment on the $\mu$-term and $B$-term.

\subsection{Numerical analyses}

As discussed in the introduction, the messenger fields generally do
not form complete $SU(5)$ multiplets as far as they preserve the
unification of the standard model gauge couplings. The $SU(2)$ doublet
and $SU(3)$ triplet parts of the messengers can have different
supersymmetric masses and SUSY breaking mass splittings. We here
consider this general situation and take the different couplings 
${c_P^2}_{ij}, {c_P^3}_{ij}$ and ${\lambda_P^2}_{ij},
{\lambda_P^3}_{ij}$ 
for the doublet and  triplet parts of the messenger fields to obtain the general spectra of the visible sector fields. Then, we have five continuous parameters $F_\phi, \Lambda_g^2, \Lambda_g^3, \Lambda_X^2, \Lambda_X^3$ defined as follows:
\begin{equation}
\begin{split}
\Lambda_g^{2,3} &= \frac{\del}{\del X} \log{\text{det}\mathcal{M}^{2,3}} \Big|_{X= \langle X \rangle} F_{X}, \\
(\Lambda_X^{2,3})^2 &= \sum_i \left( \frac{\del \log{|a_i^{2,3}|}}{\del \log{|X|}} \right)^2 \left| \frac{F_{X}}{X} \right|^2 + \frac{\del}{\del \log{|X|}} \log{|\text{det}\mathcal{M}^{2,3}|} F_\phi \frac{F_{X}^\dagger}{{X^\dagger}} + h.c.
\end{split}
\end{equation}
where the upper indices $2,3$ represent the doublet and triplet
contributions and $a_i^{2,3}$ denote eigenvalues of the
messenger mass matrix of doublet and triplet messenger fields
$\mathcal{M}^{2,3}$. 
For simplicity, here we assume $\Lambda_g^2 = \Lambda_g^3$ and 
denote them as $\Lambda_g (= \Lambda_g^2 = \Lambda_g^3)$ to 
parametrize the gaugino masses.
It is straightforward to extend the following numerical analysis 
to the case with  $\Lambda_g^2 \neq \Lambda_g^3$.
We can set the overall scale by the scale $F_\phi$ and express
all the soft mass parameters as functions of dimensionless
parameters $r_1 \equiv \Lambda_g/\Lambda_X^2$, $r_2 \equiv
\Lambda_X^2/F_\phi$ and $r_3 \equiv \Lambda_X^3/F_\phi$.

Before numerical analysis, we comment on sum rules of the soft scalar masses.
For each of three generations, the sfermion masses 
at the messenger scale satisfy the following sum rules,
\begin{eqnarray}\label{sum-rule-1}
\tr \ (B-L)\ m^2 \ = \ 2 m^2_{\tilde{Q}} -  m^2_{\tilde{U}} -  m^2_{\tilde{D}} - 
 2m^2_{\tilde{L}} +  m^2_{\tilde{E}} =0 .
\end{eqnarray}
In addition, the following sum rule:
\begin{equation}\label{sum-rule-2}
 \tr~Ym^2 \ = \ m^2_{\tilde{Q}} - 2 m^2_{\tilde{U}} +  m^2_{\tilde{D}} - 
 m^2_{\tilde{L}} +  m^2_{\tilde{E}} = 0, 
\end{equation}
is also satisfied for each of the first and second generations.
These sum rules are the same as those derived 
in general gauge mediation
\cite{Meade:2008wd}.\footnote{
These sum rules have also been derived in the context of 
various SUSY breaking models \cite{Martin:1993ft}.}
The latter sum rule (\ref{sum-rule-2}) is violated in the third generation 
because of effects from the top Yukawa coupling.
Thus, our parameter space is different from one of
general gauge mediation.
Instead of the above sum rule, the sfermion masses in the third 
generation 
satisfy the following sum rule,
\begin{equation}\label{sum-rule-3}
m^2_{\tilde{Q}} - 2 m^2_{\tilde{U}} +  m^2_{\tilde{D}} - 
 m^2_{\tilde{L}} +  m^2_{\tilde{E}} + m^2_{H_2} - m^2_{H_1}= 0.
\end{equation}
These sum rules have corrections due to renormalization group (RG) effects 
between the messenger scale and the weak scale 
\cite{Meade:2008wd,Jaeckel:2011ma}.
However, our natural messenger scale is low such as 
${\cal O}(10)$ TeV.
Thus, such RG corrections on the sum rules  
are small.

Also, we give a comment on the gauge coupling unification.
The doublet/triplet splitting leads to corrections such as 
$\log (\mathcal{M}^{2}/\mathcal{M}^{3})$ in the gauge coupling 
unification.
We assume that $c_P^2/c_P^3={\cal O}(1)$ and
$\lambda_P^2/\lambda_P^3={\cal O}(1)$.
That would lead to $\mathcal{M}^{2}/\mathcal{M}^{3}={\cal O}(1)$.
Then, we assume that the doublet/triplet splitting would have 
a sufficiently small effect on the gauge coupling unification 
with leading to ${\cal O}(1)$ of splitting for $r_2/r_3$.

In numerical analysis, we take the
messenger scale as $10 \, \text{TeV}$. The soft SUSY breaking masses
are evaluated using the expressions of \eqref{gauginomass} and
\eqref{scalarmass}, and evolved down to the weak scale 
with RG effects. 
Here we take $\tan \beta = 10$.

One of the important constraints is the condition to 
avoid the tachyonic slepton masses, 
which always appear in the pure anomaly mediation.
The slepton masses squared become positive for 
the following parameter region,
\begin{equation}\label{eq:slepton-mass}
1.5r^2_2 + r^2_3 ~\gtrsim~ 19,
\end{equation}
at the messenger scale.
There are RG corrections due to the bino mass 
between the messenger scale and the weak scale, 
but such corrections are small and the allowed region 
does not change drastically.

In Figure 1, we show the spectrum at the weak scale with $r_2=r_3$. The
upper panel corresponds to the case with $r_1=0$, where the gaugino
masses are induced only by the pure anomaly mediation. 
The lower panel corresponds to the case with $r_1=1$, that is,  both
anomaly mediation and gauge mediation contribute to the
gaugino masses. The red, green and blue solid lines represent the
bino, wino and gluino, respectively. The pink dot, yellow dash, violet
dashdot, brown longdash, gold spacedash and black spacedot lines
represent the soft masses of the left-handed stop $\tilde{Q}_3$, 
the right-handed stop $\tilde{U}_3$, the right-handed down-sector 
squarks $\tilde{D}$, the left-handed sleptons $\tilde{L}$, 
the right-handed sleptons $\tilde{E}$ and the up-sector Higgs $H_2$, 
where the soft mass of the down-sector Higgs $H_1$ is 
the same as one of the left-handed sleptons $\tilde{L}$. 
In Figure 1, we have omitted masses of the first two generations of 
the left-handed squarks $\tilde{Q}_{1,2}$ and the right-handed 
up-sector squarks $\tilde{U}_{1,2}$.
Those are heavier than  $\tilde{Q}_3$ and $\tilde{U}_3$.
In Figures 2, 3 and 4, we take $r_2 = 2
r_3, 5r_3, 1/2r_3$, the others are the same as Figure 1.

At the upper panels of Figures, 1, 2, 3 and 4, 
the gaugino masses are obtained as the pure anomaly mediation, 
because of $r_1=0$.
That is, the wino is the lightest among the gaugino fields, 
and the gluino is much heavier.
Thus, the LSP is the wino-like neutralino in the allowed region, 
where the slepton masses squared are positive (\ref{eq:slepton-mass}).
The next-to-LSP is the chargino and the mass difference between 
the LSP and the next-to-LSP is small like anomaly mediation.
Squarks are much heavier than other sparticles in the allowed regions.
Obviously, a large value of $r_3$ like $r_2=r_3/2$ in the upper panel
of Figure 4 makes 
squarks heavier and a small value like $r_2=2 r_3$ and $r_2=5 r_3$ 
in the upper panels of Figures 2 and 3 make them lighter.
The soft scalar mass of $H_2$ behaves non-trivially 
depending of $r_2$ and $r_3$.
For $r_2=r_3$ and $r_2=r_3/2$ in the upper panels of 
Figures 1 and 4,
the soft scalar mass squared $m_{H_2}^2$ decreases as 
$|r_2|$ increases.
For $r_2 = 2r_3$ in the upper panel of Figures 2,  
the soft scalar mass $m_{H_2}$ is almost constant 
against $r_2$.
On the other hand, for $r_2 = 5r_3$ in the upper panel of Figures 3,  
the soft scalar mass squared $m_{H_2}^2$ increases as 
$|r_2|$ increases.
The soft scalar mass squared $m_{H_2}^2$ becomes positive 
for $|r_2| \gtrsim 7.4$, while it is always negative in 
the other figures.
This non-trivial behavior is originated from the 
negative radiative corrections due to the stop masses 
between the  messenger scale and the weak scale.
The stop masses are quite heavy in the upper panels of 
Figures 1 and 4.
In particular, they become much heavier as $|r_2|$ increases.
They lead to largely negative radiative corrections
in $m_{H_2}^2$.
In the case with $r_2=2 r_3$ and $r_2=5 r_3$, 
the stop masses are not heavy compared with the above cases.
In addition, the gauge mediation effect on 
$m_{H_2}^2$ is positive and it increases as $|r_2|$ increases.
This leads to the behaviors shown in the upper panels of 
Figures 2 and 3.
For $r_2 \gtrsim 2r_3 $, there appears the parameter region 
of $r_2$, where $m_{H_2}^2$ is positive at the weak scale.
In such a region, the successful electroweak symmetry 
breaking does not occur.
Then, we have the excluded region for a large $|r_2|$.

In the lower panels of Figures 1, 2, 3 and 4, 
it seems that several masses vary variously 
depending on a value of $r_2$ as well as $r_3$.
Such a behavior is originated from the fact that 
for $r_1\neq 0$ the gaugino masses have contributions due to 
both anomaly mediation and gauge mediation, 
and they vary depend on $r_1r_2$ in our parametrization.
Varying the gaugino masses also affect behaviors of 
the scalar masses through radiative corrections between 
the messenger scale and the weak scale.
However, such radiative corrections on the slepton masses are very 
small and the behaviors of the sleptons in 
the lower panels of Figures 1, 2, 3 and 4 are almost the same as 
the corresponding upper panels, 
although the squark masses and the up-sector Higgs soft mass 
change significantly.
That is, the region for the non-tachyonic slepton masses corresponds to 
Eq.~(\ref{eq:slepton-mass}).
The three gaugino masses, $M_1$,$M_2$ and $M_3$, are proportional to 
$|r_1 r_2 - 33/5|$, $|r_1 r_2 - 1|$ and $|r_1 r_2 + 3|$, 
respectively, up to the gauge couplings.
As a result, the gaugino masses, $M_1$, $M_2$ and $M_3$, are 
very suppressed around $r_1 r_2 \approx 33/5$, $r_1 r_2 \approx 1$ 
and $r_1 r_2 \approx -3$, respectively.
On the other hand, far away from those points, 
the gaugino masses, in particular the gluino mass, 
become heavier.
For example, for $r_1=1$,  those points correspond to 
$r_2 = 33/5$, 1 and $-3$, respectively.
Because of this behavior, the wino is heavier than 
the right-handed slepton in most of the parameter space, 
and the wino can not be the LSP.
When we take a large value of $|r_1|$ like $|r_1| \gtrsim 2$, 
we would have the parameter region with the wino LSP.
For $r_2 > 0$, the bino is the LSP except the parameter 
region, where the slepton has a tachyonic mass or the 
right-handed slepton is the LSP.
On the other hand, for $r_2 < 0$, the gluino can be the LSP 
in a narrow region, but in the other region 
the right-handed slepton is the LSP.
Such a parameter region would be unfavorable.
In addition, far away from the point $r_2= -3$, the gluino 
becomes heavier.
One can see this behavior by comparing the upper and 
lower panels in Figures 1, 2, 3 and 4.

We comment on the squark masses in the lower panels of Figures 1, 2, 3 and 4.
The radiative corrections due to the gluino mass 
between the messenger scale and the weak scale are important in the squark masses.
Since the gluino mass is smaller for $r_2 < 0$ than one 
for  $r_2 > 0$, the squark masses are also smaller for $r_2 < 0$ 
than those for  $r_2 > 0$.
Note that for $r_1=0$ the squark masses as well as the slepton masses 
are symmetric under the $Z_2$ reflection $r_2 \leftrightarrow - r_2$.
Furthermore, a small value of $r_3$ like $r_2= 2 r_3$ and $r_2= 5 r_3$ 
 also leads to smaller squark masses as the upper panels of Figures 
2 and 3.
In particular, the right-handed stop can have a tachyonic mass 
in a certain parameter region, as pointed out 
already in Ref.~\cite{Hsieh:2006ig}.
Such a parameter region with the tachyonic right-handed stop 
becomes wider as $r_3$ becomes smaller like 
$r_2= 2 r_3$ and $r_2= 5 r_3$ in the lower panels of 
Figures 2 and 3.
Thus, in these cases, there is a wide region excluded by 
tachyonic masses of the right-handed slepton 
and the right-handed stop.
On the other hand, since far away from the point $r_2 = -3$, 
the gluino mass becomes heavier in particular for positive $r_2$, 
the stop masses also become heavier in these parameter region.
One can see this behavior by comparing the stop and gluino masses 
in the upper and lower panels of Figures 1, 2, 3 and 4.

Here, we comment on the up-sector Higgs soft mass in the lower panels of Figures 1, 2, 3 and 4.
The behavior of the up-sector Higgs soft mass depends 
on the stop masses.
When the stop masses of the lower panels of Figures 1, 2, 3 and 4
are similar to those in the corresponding upper panels, 
values of $m_{H_2}^2$ are similar between the upper and lower 
panels.
When the stop masses of the lower panels become heavier
than those in the upper panels,  a value of $m_{H_2}$ is 
driven to a negative direction in the lower panels 
compared with those in the upper panels.
For a large value of $|r_2|$ like $r_2=5r_3$, 
there is a parameter space with $m^2_{H_2} > 0$ in 
the lower panel of Figure 3, similar to the 
corresponding upper panel.
Thus, the parameter region is constrained by 
realization of the successful electroweak symmetry 
breaking as well as avoiding the tachyonic 
slepton and/or stop masses.
Indeed, the allowed region corresponds to   $4 \lesssim r_2 \lesssim
15$ in the lower panel of Figure 3.
To summarize phenomenological aspects shown 
in the lower panels of Figures 1, 2, 3 and 4, 
certain parameter regions are excluded by the masses 
of the right-handed slepton and/or the right-handed stop.
In the half of the allowed region, i.e. $r_2 >0$, 
the bino is the LSP, while in the other half $r_2 < 0$ 
the right-handed slepton would be the LSP except 
a narrow region with the gluino LSP.

When we simply compare between 
the cases with $r_1 = 0$ and $r_1=1$, 
the allowed region for $r_1 =0$ is wider than one for $r_1=1$.
If we take different values of $r_1$, the situation would change.

We give several representative points in Tables 1 and 2. We take 
$F_\phi$ i.e. the gravitino mass to satisfy the wino mass bound
$m_{\tilde{W}} > 100 \text{GeV}$ and the bino mass bound
$m_{\tilde{B}} > 50 \text{GeV}$. 
The $A$-term is induced by the pure anomaly mediation.
We also estimate the $A$-term
corresponding to the top Yukawa coupling. It is found that $A_t$ is smaller than the stop mass in most parameter space, so the stop mixing is relatively-small.
At all of the points shown in Tables 1 and 2, 
the stop masses are heavy, so that the lightest Higgs mass is heavy 
enough to satisfy the LEP bound $m_h > 114$ GeV.

In the model, which we discussed in section 2, the natural messenger 
scale would be of ${\cal O}(10)$-${\cal O}(100)$ TeV.
However, other types of models would lead to a mixture 
between anomaly mediation and gauge mediation with different 
messenger scales \cite{Pomarol:1999ie}.
Following such a rather phenomenological viewpoint, finally 
we show examples with higher messenger scales.
Figure 5 shows the soft masses in the case with $r_1=0$ and $r_2 =r_3$, 
which are the same as the upper panel in Figure 1.
The upper and lower panels in Figure 5 correspond to 
$10^{10}$ GeV and $2 \times 10^{16}$ GeV as the messenger scale.
It seems that 
the qualitative results except the up-sector Higgs soft mass 
are roughly similar to the case that the messenger scale is 10 TeV.
One of the important differences is that the spectrum of 
the soft scalar masses are rather compact in the cases with 
higher messenger scales compared with the case of the 10 TeV messenger
scale, that is, the squarks are lighter in particular for a large 
value of $|r_2|$.
That also affects on the soft scalar mass of the up-sector Higgs.
Because of the lighter stop masses, the soft mass squared 
of the up-sector Higgs becomes positive for a large value 
of $|r_2|$.
This situation is similar to one in the upper panel of Figure 3.
Another important point is the long logarithmic RG running.
In particular, the slepton masses receive such a long logarithmic RG
running 
effect due to the bino mass, and they tend to become positive.
Thus, the region excluded by the tachyonic slepton 
becomes narrow in Figure 5 compared with one 
in the upper panel of Figure 1.
Similarly, we can study other values of $r_1$ and $r_3$ 
for higher messenger scales for a purely phenomenological purpose.

We have taken $\Lambda^2_g = \Lambda^3_g$ just for simplicity 
in all of the above analyses.
It would be interesting to study for other values of 
the ratio, $\Lambda^2_g / \Lambda^3_g \neq 1$.

\subsection{The $\mu-B\mu$ problem}

Here, we comment on the $\mu$ term and the $B\mu$ term. 
In our model, it would be simple to generate the $\mu$ term and the
$B\mu$ term by the following terms,
\begin{equation}
\int d^4 \theta c_H \frac{\phi^\dagger}{\phi} H_1 H_2 ~~+~ 
\int d^2 \theta \lambda_H S H_1 H_2 ~~ +~~  h.c.
\end{equation}
Then, we obtain 
\begin{equation}
\begin{split}
\mu &= \fphi c_{H} + \vs \lambda_{H}, \\
B\mu &= -\fphi \mu + \left(\fphi \vs + \fs \right) \lambda_{H}.
\end{split}
\end{equation}
These are independent of each other because there are two parameters, 
$c_H$ and $\lambda_H$.
However, the natural scale of $B$ would be of ${\cal O}(F_\phi)$, 
and such a scale is too large to realize successfully the electroweak 
symmetry breaking.
For example, when $|m^2_{H_1}| \sim |m^2_{H_2}| $, 
it would be required that $\mu^2$ and $B \mu $ are of the same order.
Such a value of $\mu$ can be obtained for 
$c_H, \lambda_H = \mathcal{O}(0.1-0.01)$ 
since we assume that $\fphi, \vs  = \mathcal{O}(10) \, \text{TeV}$. 
However, since the natural scale of $B$ would be of
$\mathcal{O}(F_\phi)$, 
we would need a few percent of fine-tuning between $c_H$ and
$\lambda_H$ to realize 
\begin{equation}
\mu B \sim \mu^2, ~|m^2_{H_1}|,  ~|m^2_{H_2}| ,
\end{equation}
that is, the $\mu-B\mu$ problem.
Each of anomaly mediation and gauge mediation 
has the $\mu-B\mu$ problem.
Obviously, their mixture studied here also has the same 
problem unless we have any definite mechanism to cancel 
out the contributions due to anomaly mediation 
and gauge mediation in the $B\mu$ term to lead to a 
suppressed value of the $B \mu$ term.\footnote{
For example, for the mixture of anomaly mediation 
and moduli mediation, there is a certain type of 
cancellation mechanisms in the $B\mu$ term \cite{Choi:2005hd}. 
We need such a cancellation mechanism in this scenario.}

On the other hand, in Figure 3, there is a parameter region, where 
$|m_{H_2}^2|\ll  |m_{H_1}^2|$, e.g. Point 4 in Table 1. 
When the relation
\begin{equation}
\mu^2 \sim m_{H_2}^2 \ll B\mu \ll m_{H_1}^2 ,
\end{equation}
is satisfied, we can realize the successful electroweak symmetry 
breaking \cite{Csaki:2008sr,Choi:2006xb,Kobayashi:2009rn}, 
that is, a large value of $B$ may not be problematic.
 For example, at Point 4, we have
\begin{equation}
m_{H_1}^2 (M_Z) \simeq 1.61~ \text{TeV}^2, 
\quad m_{H_2}^2 (M_Z) \simeq - 7870~ \text{GeV}^2.
\end{equation}
By using
\begin{equation}
|\mu|^2 = -\frac{M_Z^2}{2} -
\frac{m_{H_2}^2\tan^2{\beta}-m_{H_1}^2}{\tan^2{\beta}-1}, \quad  \sin{2\beta} = \frac{2B \mu}{2|\mu|^2 + m_{H_1}^2 + m_{H_2}^2},
\end{equation}
we obtain
\begin{equation}
|\mu| \simeq 120~ \text{GeV}, \quad  B \simeq 1.4~ \text{TeV},
\end{equation}
e.g. for $\tan{\beta}=10$.
In this case, the fine-tuning to be required between 
$c_H$ and $\lambda_H$ is ameliorated such as ${\cal O}(10)$ \%.
However, we need another fine-tuning for $r_2$ to realize 
$|m_{H_2}^2|\ll  |m_{H_1}^2|$, 
unless we have any definite mechanism to set a proper value of 
$r_2$.
Thus, the simple way to generate the $\mu$ term and 
the $B \mu$ term requires a fine-tuning.
We could consider another way to generate the $\mu$ term and 
the $B \mu$ term \cite{Hsieh:2006ig,Cai:2010tj}, 
where we do not need fine-tuning, but 
here we do not pursue further.
Finally, we comment on the LSP.
One of specific aspects at Point 4 is that the LSP is 
higgsino-like,  while different points lead to 
another LSP such as the bino or wino.


\begin{table}[htbp]
\begin{center}
\begin{tabular}{|c|c|c|c|c|c|}
  \hline
   & Point 1 & Point 2 & Point 3 & Point 4 & Point 5  \\
  \hline
  $r_1$ & 0 & 0 & 0 & 0 & 0  \\
  \hline
  $r_2$ & 3.5 & 4.0 & 4.0 & 7.37 & 2.5 \\
  \hline
  $r_3$ & $r_2$ & $r_2/2$ & $r_2/5$ & $r_2/5$ & $2r_2$  \\
  \hline
  $m_{3/2}$ & $50 \, \text{TeV}$ & $50 \, \text{TeV}$ & $50 \, \text{TeV}$ & $50 \, \text{TeV}$ & $50 \, \text{TeV}$  \\
  \hline
  $m_{\tilde{B}}$ & 462 & 462 & 462 & 462 & 462 \\
  \hline
  $m_{\tilde{W}}$ & 135 & 135 & 135 & 135 & 135 \\
  \hline  
  $m_{\tilde{G}}$ & 1430 & 1430 & 1430 & 1430 & 1430 \\
  \hline 
  $m_{\tilde{Q}_3}$ & 2450 & 1800 & 1450 & 1910 & 3200 \\
  \hline
  $m_{\tilde{U}_3}$ & 1970 & 1130 & 465 & 753 & 2810 \\
  \hline 
  $m_{\tilde{Q}_{1, 2}}$ & 2590 & 1940 & 1580 & 2040 & 3370 \\
  \hline
  $m_{\tilde{U}_{1, 2}}$ & 2302 & 1480 & 970 & 1260 & 3180 \\  
  \hline
  $m_{\tilde{D}}$ & 2530 & 1820 & 1430 & 1624 & 3340 \\
  \hline
  $m_{\tilde{L}} (m_{H_1})$ & 586 & 163 & 671 & 1268 & 412 \\
  \hline
  $m_{\tilde{E}}$ & 209 & 192 & 166 & 431 & 232 \\
  \hline
  $-m_{H_2}$ & 1375 & 1020 & 856 & 47.5 & 1790 \\
  \hline
  $A_t$ & 1300 & 1300 & 1300 & 1300 & 1300 \\
  \hline
\end{tabular}
\end{center}
\caption{We give various representative points at the weak scale in
  the parameter space with $r_1=0$. For the gaugino masses we take $|M|$, and
  for the scalar masses we take $|m^2|^{1/2} \times
  \text{sign}(m^2)$. The messenger scale is taken to be $10$ TeV. All
  masses except $m_{3/2}$ are in GeV.}
\end{table}

\begin{table}[htbp]
\begin{center}
\begin{tabular}{|c|c|c|c|c|c|}
  \hline
   & Point 6 & Point 7 & Point 8 & Point 9 & Point 10 \\
  \hline
  $r_1$ & 1 & 1 & 1 & 1 & 1 \\
  \hline
  $r_2$ & 4.0 & 4.5 & 4.5 & 2.5 & 3.5  \\
  \hline
  $r_3$ & $r_2$ & $r_2/2$ & $r_2/5$ & $2r_2$ & $2r_2$  \\
  \hline
  $m_{3/2}$ & $15\, \text{TeV}$ & $20\, \text{TeV}$ & $20 \, \text{TeV}$ & $25 \, \text{TeV}$ & $15 \, \text{TeV}$ \\
  \hline
  $m_{\tilde{B}}$ & 54.6 & 58.8 & 58.8 & 144 & 65.1  \\
  \hline
  $m_{\tilde{W}}$ & 122 & 189 & 189 & 101 & 101  \\
  \hline  
  $m_{\tilde{G}}$ & 998 & 1430 & 1425 & 1306 & 926  \\
  \hline 
  $m_{\tilde{Q}_3}$ & 961 & 1070 & 954 & 1713 & 1376  \\
  \hline
  $m_{\tilde{U}_3}$ & 835 & 875 & 743 & 1527 & 1260  \\
  \hline 
  $m_{\tilde{Q}_{1, 2}}$ & 1010 & 870 & 1000 & 1800 & 1440 \\
  \hline
  $m_{\tilde{U}_{1, 2}}$ & 936 & 994 & 856 & 1710 & 1400 \\  
  \hline
  $m_{\tilde{D}}$ & 987 & 1080 & 954 & 1788 & 1430 \\
  \hline
  $m_{\tilde{L}} (m_{H_1})$ & 207 & 310 & 671 & 207 & 185  \\
  \hline
  $m_{\tilde{E}}$ & 76.1 & 91.0 & 80.2 & 112 & 112  \\
  \hline
  $-m_{H_2}$ & 487 & 507 & 441 & 932 & 716   \\
  \hline
  $A_t$ & 590 & 819 & 819 & 858 & 564  \\
  \hline
\end{tabular}
\end{center}
\caption{Same as Table 1, but for $r_1=1$.}
\end{table}

\section{Conclusion}
We have studied the models that anomaly mediation and gauge mediation
are competed. 
This mixture can avoid the tachyonic sleptons in a certain 
parameter space.
Our messenger structure, (\ref{model}) and (\ref{model2}), 
leads to a quite rich pattern of the SUSY breaking terms.
Still, there are the parameter regions excluded by 
the tachyonic sleptons, the tachyonic stops or the 
positive soft scalar mass of the up-sector Higgs field.
It seems that the allowed parameter space for $r_1=0$ 
is wider than one for $r_1=1$.
Thus, the models, where the gaugino masses are 
generated by the pure anomaly mediation, 
would be interesting.
Obviously, our models naturally solve the SUSY flavor problem, 
because of the mixture between gauge mediation and 
anomaly meditation.
The LSP can be bino-like, wino-like or higgsino-like 
depending on the parameters, while the LSP might 
be the stau or the stop in a certain region.
Thus, we have a dark matter candidate as usual 
and it would be interesting to study dark matter physics 
in our models.
The gravitino is heavy such as 
$\fphi \sim \mathcal{O}(10) \, \text{TeV}$.

We have studied the model with the $5 + \bar 5$ messenger fields.
We can extend our models by adding $10 + \bar {10}$ 
messenger fields.
When we split $10$ into $(3,2) + (\bar 3,1) + (1,1)$ \cite{Buican:2008ws}, 
our model would cover the parameter space corresponding to 
the general gauge mediation \cite{Meade:2008wd} 
with anomaly mediation.
Those models can lead to much richer structure 
of the SUSY breaking terms.

We have derived the sum rules among the soft scalar masses.
The sum rules for the first and the second generations are the same as those in general 
gauge mediation, but the third generation leads to 
the different sum rule.
Thus, our parameter space is different from one 
in general gauge mediation.
It is important to study theoretical implications 
and phenomenological aspects of our sum rules.

\section*{Acknowledgments}

We would like to thank T. Shimomura for useful discussions. T.K. is supported in part by the Grant-in-Aid for Scientific Research No. 20540266 and the
Grant-in-Aid for the Global COE Program ``The Next Generation of Physics, Spun from
Universality and Emergence'' from the Ministry of Education, Culture,Sports, Science
and Technology of Japan.

\setcounter{equation}{0}
\renewcommand{\theequation}{A.\arabic{equation}}

\section*{A: Summary of the soft parameters}
Here, we show the explicit formula of the soft masses in our scenario.
The group-theoretical factors $C_2$ and $a$ are summarized in Table 3.
\begin{table}[t]
\begin{center}
{\renewcommand\arraystretch{2.5}
\begin{tabular}{|c|c|c|c|c|c|c|c|}
   \hline
     & $Q$ & $U$ & $D$ & $L$ & $E$ & $H_1$ & $H_2$ \\
   \hline
   $C_2^1$ & $\displaystyle{\frac{1}{60}}$ & $\displaystyle{\frac{4}{15}}$ & $\displaystyle{\frac{1}{15}}$ & $\displaystyle{\frac{3}{20}}$ & $\displaystyle{\frac{3}{5}}$ & $\displaystyle{\frac{3}{20}}$ & $\displaystyle{\frac{3}{20}}$ \\
   \hline
   $C_2^2$ & $\displaystyle{\frac{3}{4}}$ & $0$ & $0$ & $\displaystyle{\frac{3}{4}}$ & $0$ & $\displaystyle{\frac{3}{4}}$ & $\displaystyle{\frac{3}{4}}$ \\
   \hline
   $C_2^3$ & $\displaystyle{\frac{4}{3}}$ & $\displaystyle{\frac{4}{3}}$ & $\displaystyle{\frac{4}{3}}$ & $0$ & $0$ & $0$ & $0$ \\
   \hline
   $a$ & $2$ & $4$ & $0$ & $0$ & $0$ & $0$ & $6$ \\
   \hline
\end{tabular}
}
\end{center}
\caption{The factors $C_2$ and $a$.}
\end{table}
The other coefficients are obtained as 
\begin{equation}
\begin{split}
b_1 = - \frac{33}{5}, \ b_2 = -1, \ b_3 = 3, \\
f^1 = \frac{13}{15}, \ f^2 = 3, \ f^3 = \frac{16}{3}, \ e = 6.
\end{split}
\end{equation}
In addition, we define 
\begin{equation}
\tilde{\alpha}_i = \left( \frac{g_i}{4\pi} \right)^2, Y_t = \left(
  \frac{y_t}{4\pi} \right)^2 .
\end{equation}

At the messenger scale, the explicit formula of the gaugino masses 
and the soft scalar masses are obtained as follows.
The gaugino masses are written by 
\begin{equation}
\begin{split}
M_1 &= \tilde{\alpha}_1 \left[ -\frac{33}{5}F_\phi + \left( \frac{3}{5} \Lambda_g^2 + \frac{2}{5}\Lambda_g^3 \right) \right], \\
M_2 &= \tilde{\alpha}_2 \left[ -F_\phi + \Lambda_g^2 \right], \\
M_3 &= \tilde{\alpha}_3 \left[ 3F_\phi + \Lambda_g^3 \right].
\end{split}
\end{equation}
The stop masses are obtained as 
\begin{equation}
\begin{split}
m_{\tilde{Q}_3}^2 &= \left[ - \frac{11}{50} \tilde{\alpha}_1^2 - \frac{3}{2} \tilde{\alpha}_2^2 + 8 \tilde{\alpha}_3^2 + Y_t^2 \left( 6Y_t^2 - \frac{13}{15} \tilde{\alpha}_1 - 3\tilde{\alpha}_2 - \frac{16}{3} \tilde{\alpha}_3 \right) \right] |F_\phi|^2 \\
&\quad + \left[ \frac{1}{30} \tilde{\alpha}_1^2 \left( \frac{2}{5} (\Lambda_X^3)^2 + \frac{3}{5} (\Lambda_X^2)^2 \right) + \frac{3}{2} \tilde{\alpha}_2^2 (\Lambda_X^2)^2 + \frac{8}{3}\tilde{\alpha}_3^2 (\Lambda_X^3)^2 \right], \\
m_{\tilde{U}_3}^2 &= \left[ - \frac{88}{25} \tilde{\alpha}_1^2 + 8 \tilde{\alpha}_3^2 + 2Y_t^2 \left( 6Y_t^2 - \frac{13}{15} \tilde{\alpha}_1 - 3\tilde{\alpha}_2 - \frac{16}{3} \tilde{\alpha}_3 \right) \right]|F_\phi|^2 \\
&\quad + \left[ \frac{8}{15} \tilde{\alpha}_1^2 \left( \frac{2}{5}
    (\Lambda_X^3)^2 + \frac{3}{5} (\Lambda_X^2)^2 \right) +
  \frac{8}{3}\tilde{\alpha}_3^2 (\Lambda_X^3)^2 \right]. \\
\end{split}
\end{equation}
The masses for the first and second generations of the up-sector
left-handed and right-handed squarks are obtained in the same form 
except taking $Y_t =0$. 
The right-handed down-sector squark masses are obtained as
\begin{equation}
\begin{split}
m_{\tilde{D}}^2 &= \left[ - \frac{22}{25} \tilde{\alpha}_1^2 + 8 \tilde{\alpha}_3^2 \right] |F_\phi|^2 + \left[ \frac{2}{15} \tilde{\alpha}_1^2 \left( \frac{2}{5} (\Lambda_X^3)^2 + \frac{3}{5} (\Lambda_X^2)^2 \right) + \frac{8}{3}\tilde{\alpha}_3^2 (\Lambda_X^3)^2 \right]. \\
\end{split}
\end{equation}
The slepton masses are obtained as 
\begin{equation}
\begin{split}
m_{\tilde{L}}^2 &= \left[ - \frac{99}{50} \tilde{\alpha}_1^2 - \frac{3}{2} \tilde{\alpha}_2^2 \right]|F_\phi|^2 + \left[ \frac{3}{10} \tilde{\alpha}_1^2 \left( \frac{2}{5} (\Lambda_X^3)^2 + \frac{3}{5} (\Lambda_X^2)^2 \right) + \frac{3}{2}\tilde{\alpha}_2^2 (\Lambda_X^2)^2 \right], \\
m_{\tilde{E}}^2 &= \left[ - \frac{198}{25} \tilde{\alpha}_1^2 \right] |F_\phi|^2 + \left[ \frac{6}{5} \tilde{\alpha}_1^2 \left( \frac{2}{5} (\Lambda_X^3)^2 + \frac{3}{5} (\Lambda_X^2)^2 \right) \right]. \\
\end{split}
\end{equation}
The Higgs soft masses are obtained as 
\begin{equation}
\begin{split}
m_{H_1}^2 &= m_{\tilde{L}}^2 \\
m_{H_2}^2 &= \left[ - \frac{99}{50} \tilde{\alpha}_1^2 - \frac{3}{2} \tilde{\alpha}_2^2 + 3Y_t^2 \left( 6Y_t^2 - \frac{13}{15} \tilde{\alpha}_1 - 3\tilde{\alpha}_2 - \frac{16}{3} \tilde{\alpha}_3 \right) \right]|F_\phi|^2 \\
&\quad + \left[ \frac{3}{10} \tilde{\alpha}_1^2 \left( \frac{2}{5} (\Lambda_X^3)^2 + \frac{3}{5} (\Lambda_X^2)^2 \right) + \frac{3}{2} \tilde{\alpha}_2^2 (\Lambda_X^2)^2 \right]. \\
\end{split}
\end{equation}


\bigskip

\appendix

\setcounter{equation}{0}
\renewcommand{\theequation}{A.\arabic{equation}}
\appendix


\bigskip

\appendix

\setcounter{equation}{0}
\renewcommand{\theequation}{A.\arabic{equation}}
\appendix

%
%

\begin{figure}[htbp]
 \begin{center}
  \begin{overpic}[width=12cm,clip]{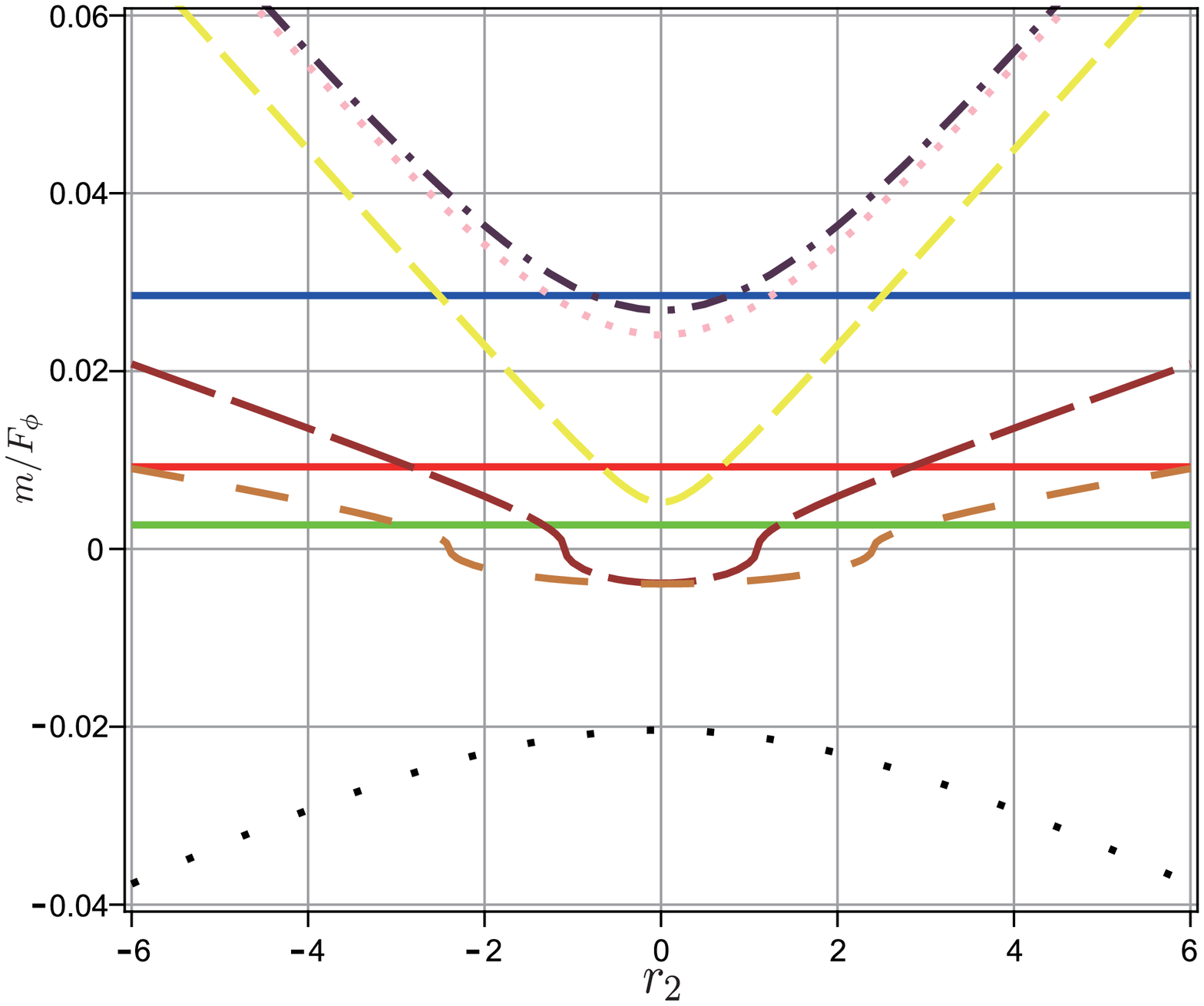}
   \put(-22,26){}
  \end{overpic}
  \begin{overpic}[width=12cm,clip]{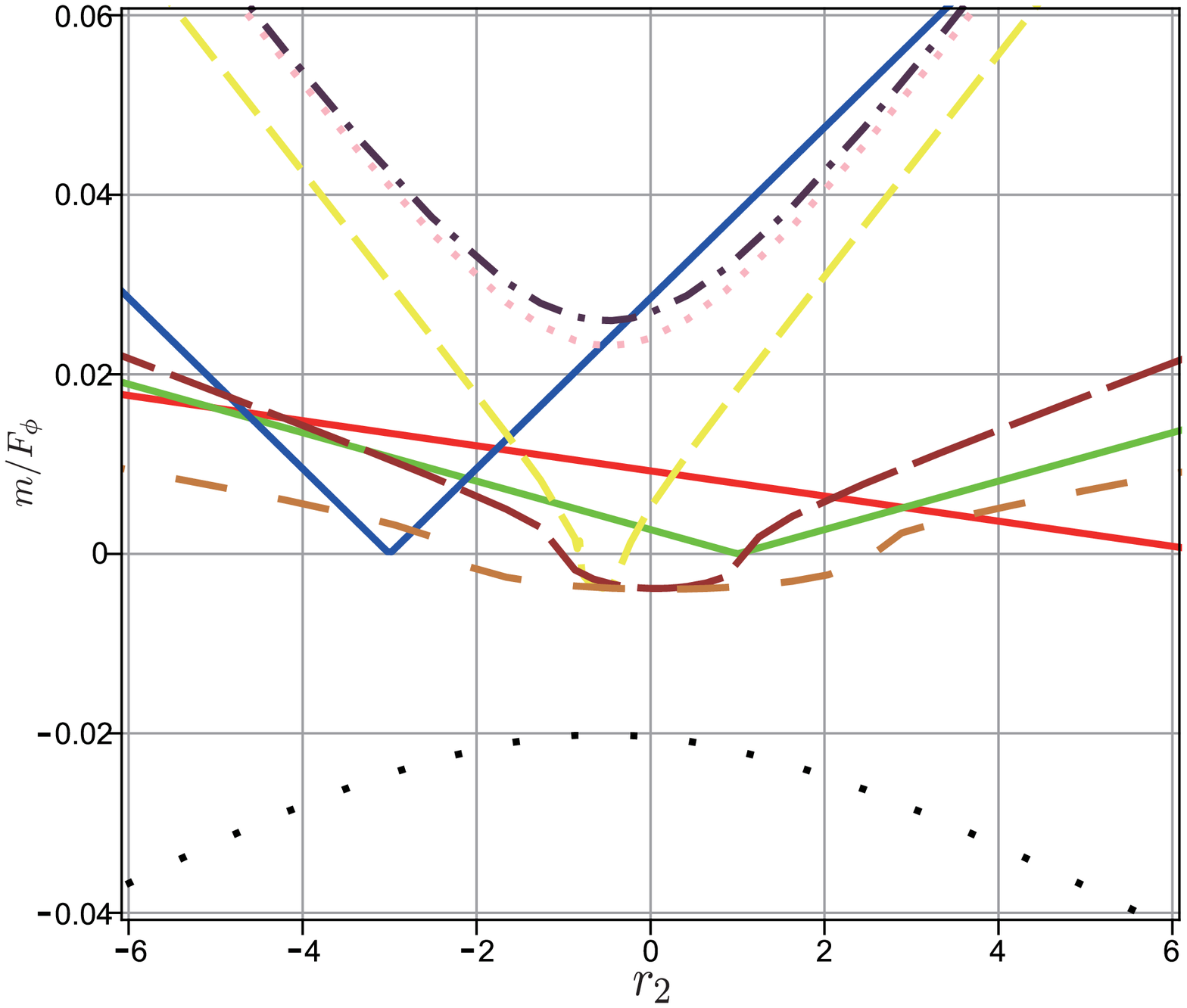}
   \put(-22,26){}
  \end{overpic}
\caption{Spectrum of the superpartner masses as a function of $r_2$. 
We take $r_2=r_3$ and the messenger scale as 10 TeV. For the gaugino masses we plot $|M|$, and for the scalar masses we plot $|m^2|^{1/2} \times \text{sign}(m^2)$ in unit of $\Lambda_\phi$.}
 \end{center}
\end{figure}

\begin{figure}[htbp]
 \begin{center}
  \begin{overpic}[width=12cm,clip]{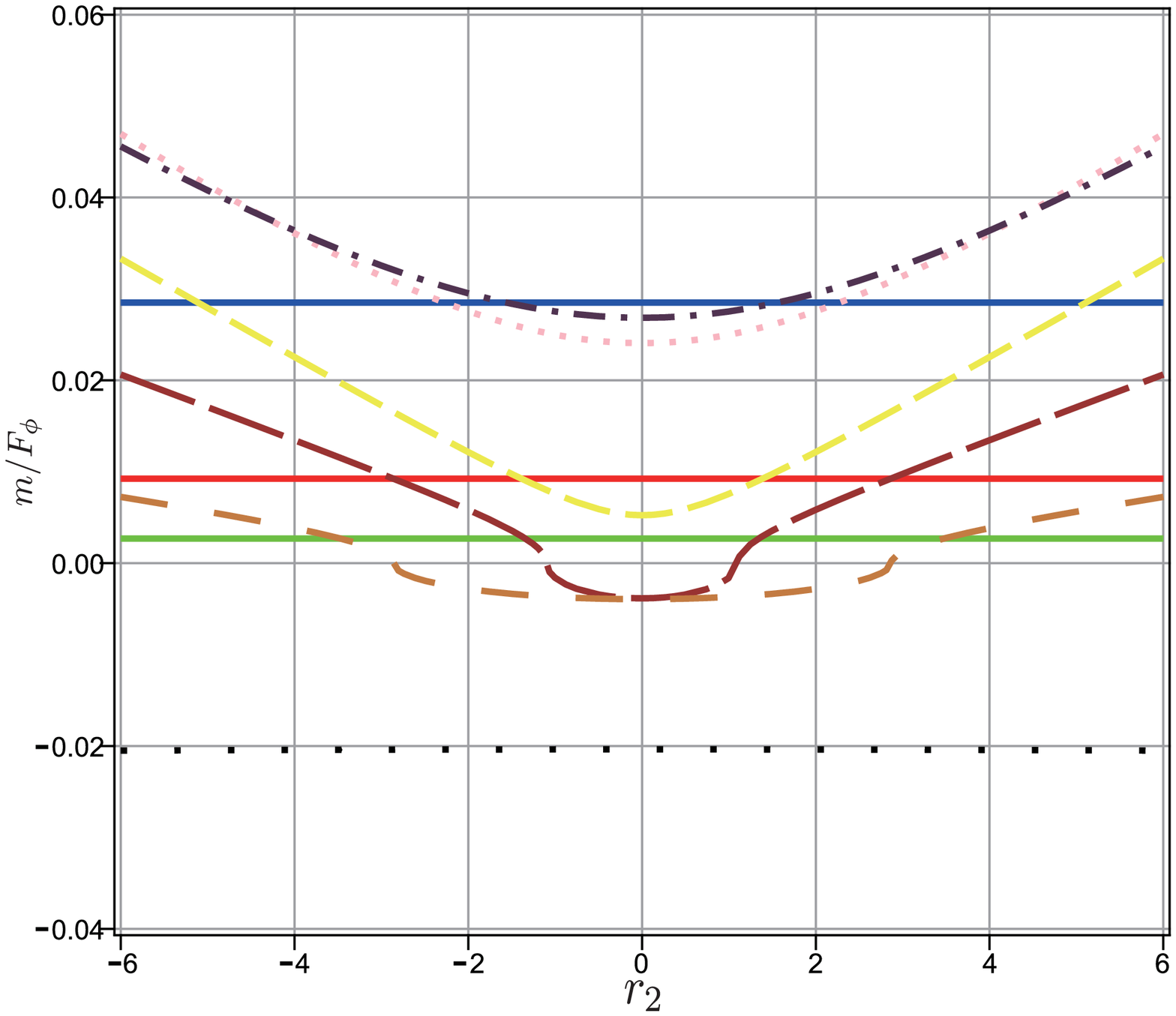}
   \put(-22,26){}
  \end{overpic}
  \begin{overpic}[width=12cm,clip]{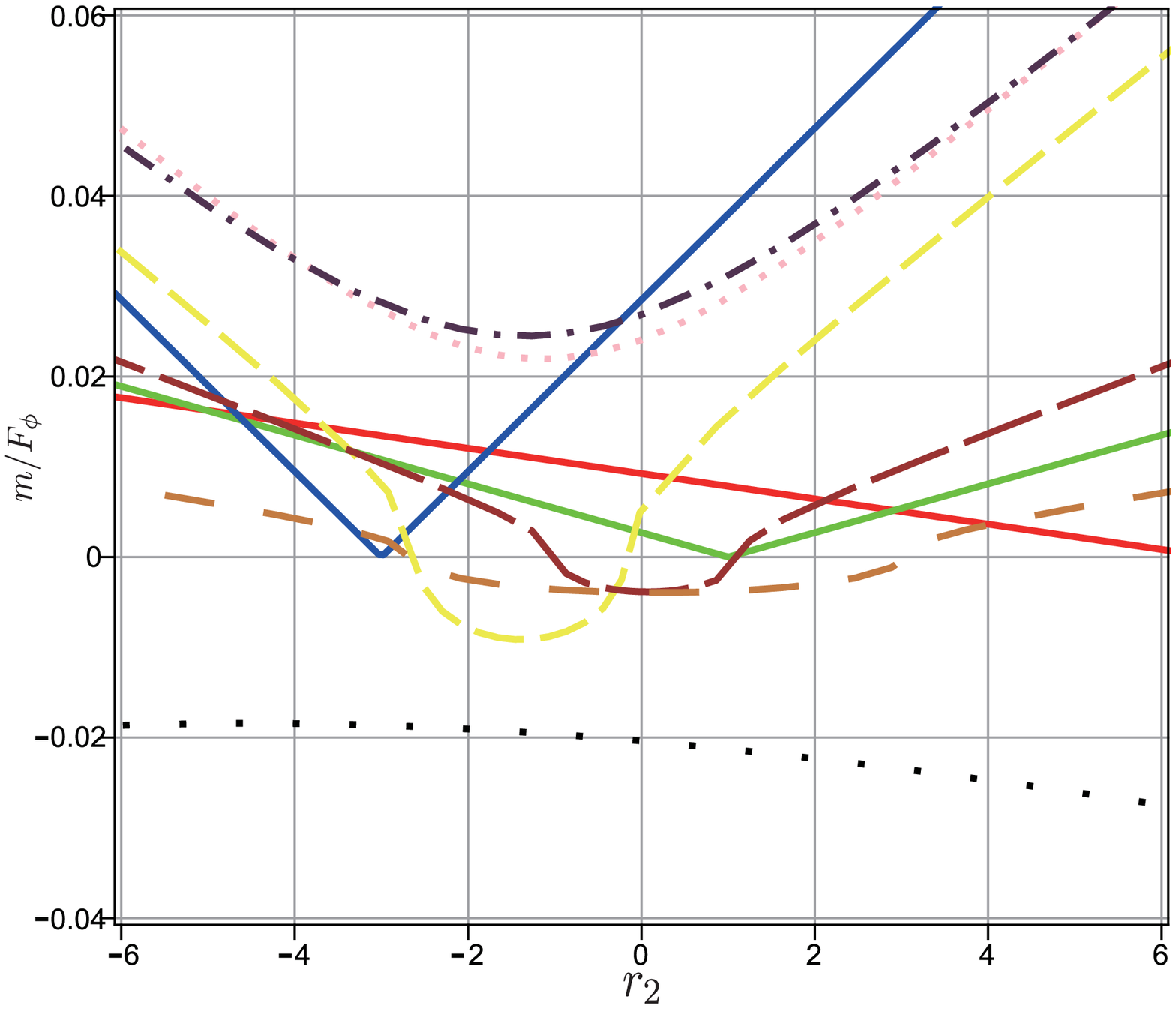}
   \put(-22,26){}
  \end{overpic}
\caption{Same as Figure 1, but for $r_2= 2r_3$.}
 \end{center}
\end{figure}

\begin{figure}[htbp]
 \begin{center}
  \begin{overpic}[width=12cm,clip]{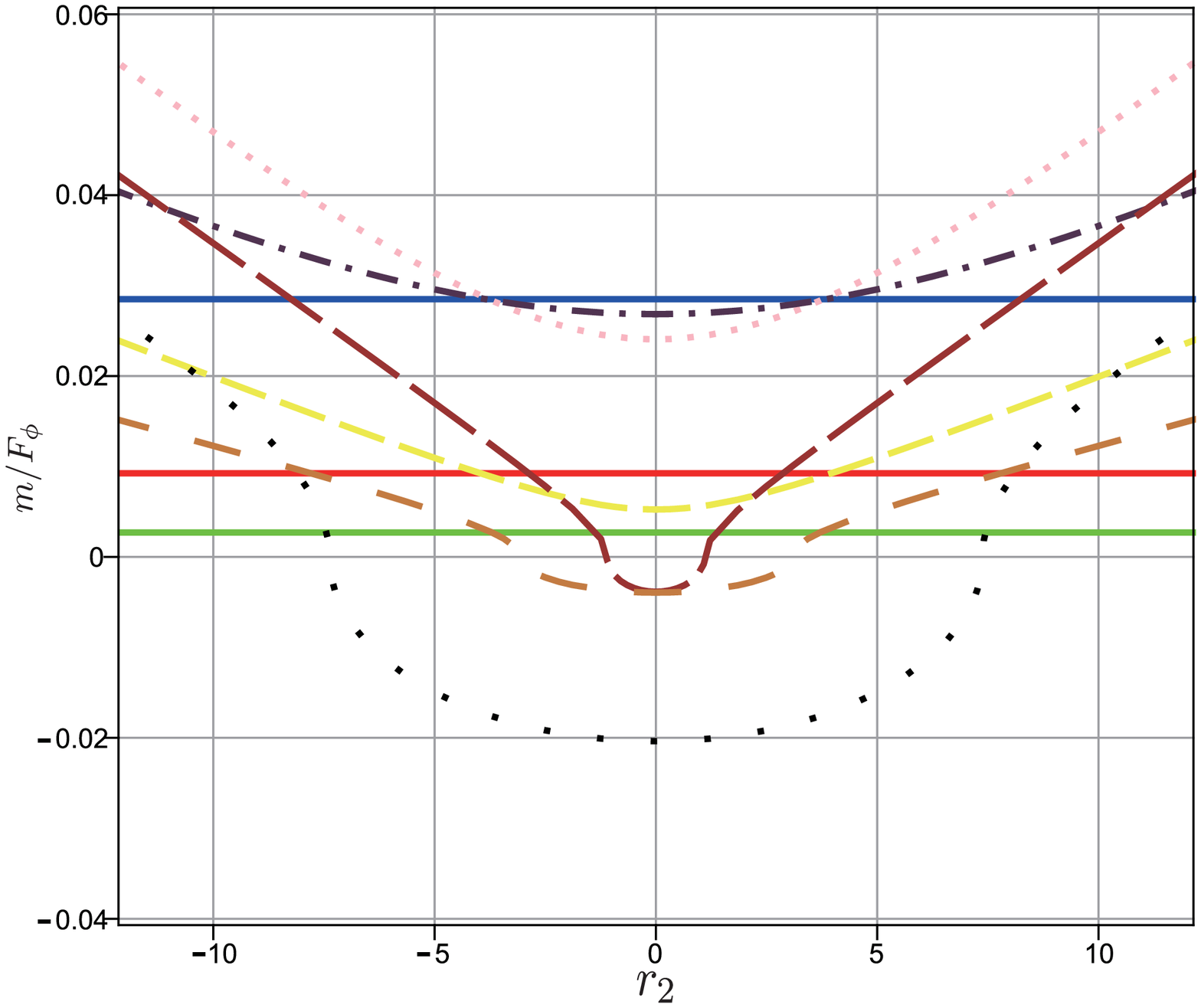}
   \put(-22,26){}
  \end{overpic}
  \begin{overpic}[width=12cm,clip]{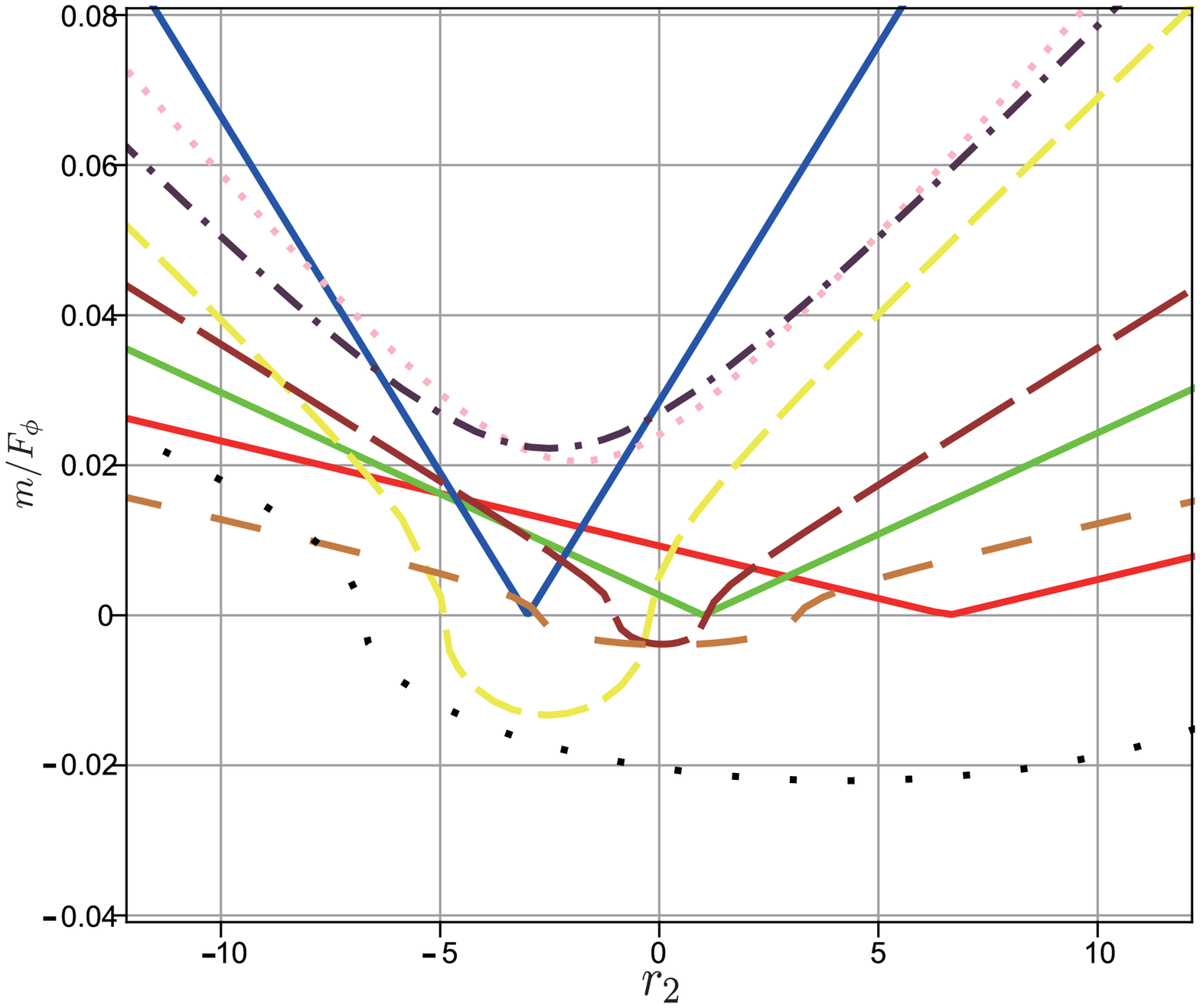}
   \put(-22,26){}
  \end{overpic}
\caption{Same as Figure 1, but for $r_2= 5r_3$.}
 \end{center}
\end{figure}

\begin{figure}[htbp]
 \begin{center}
  \begin{overpic}[width=12cm,clip]{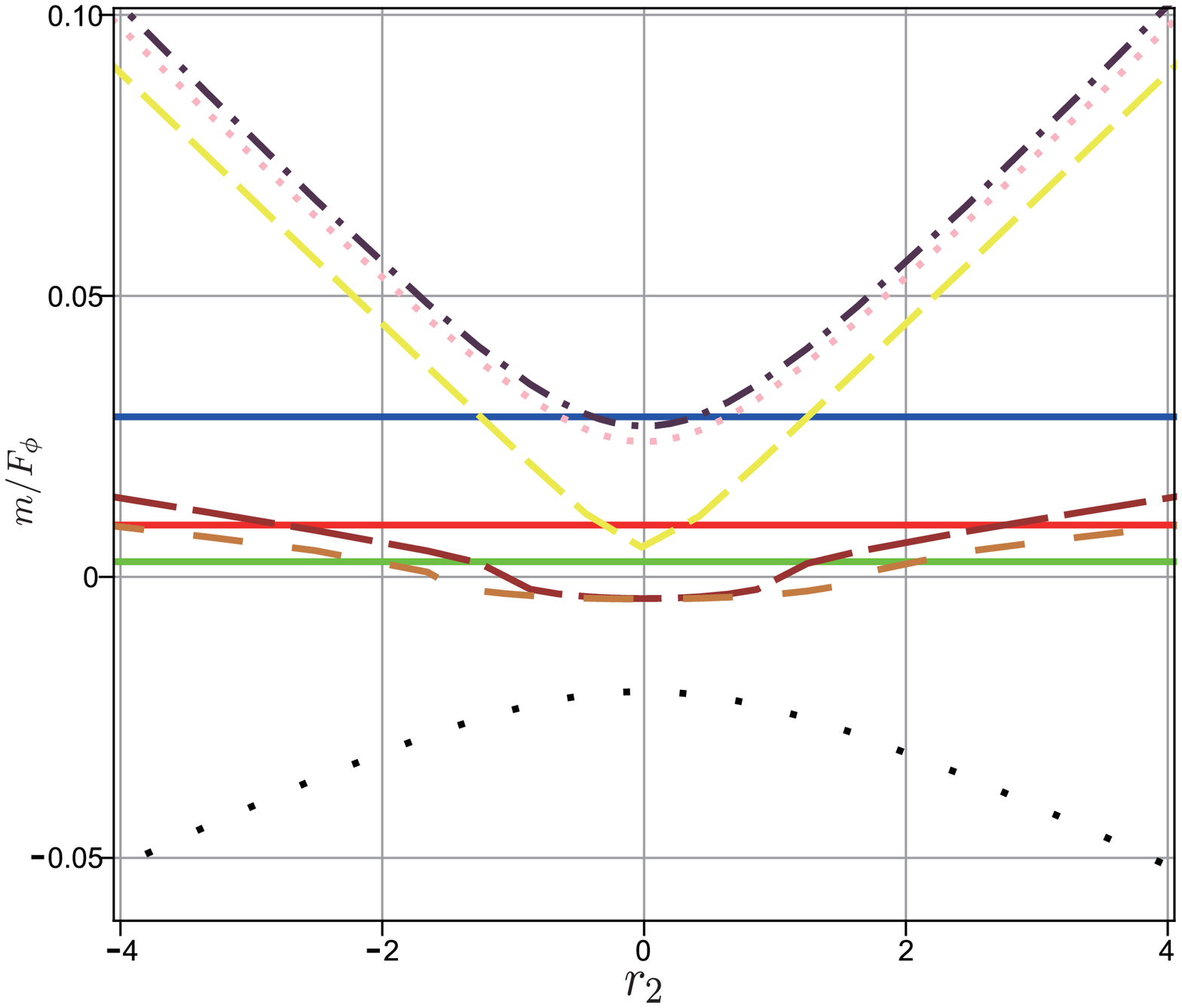}
   \put(-22,26){}
  \end{overpic}
  \begin{overpic}[width=12cm,clip]{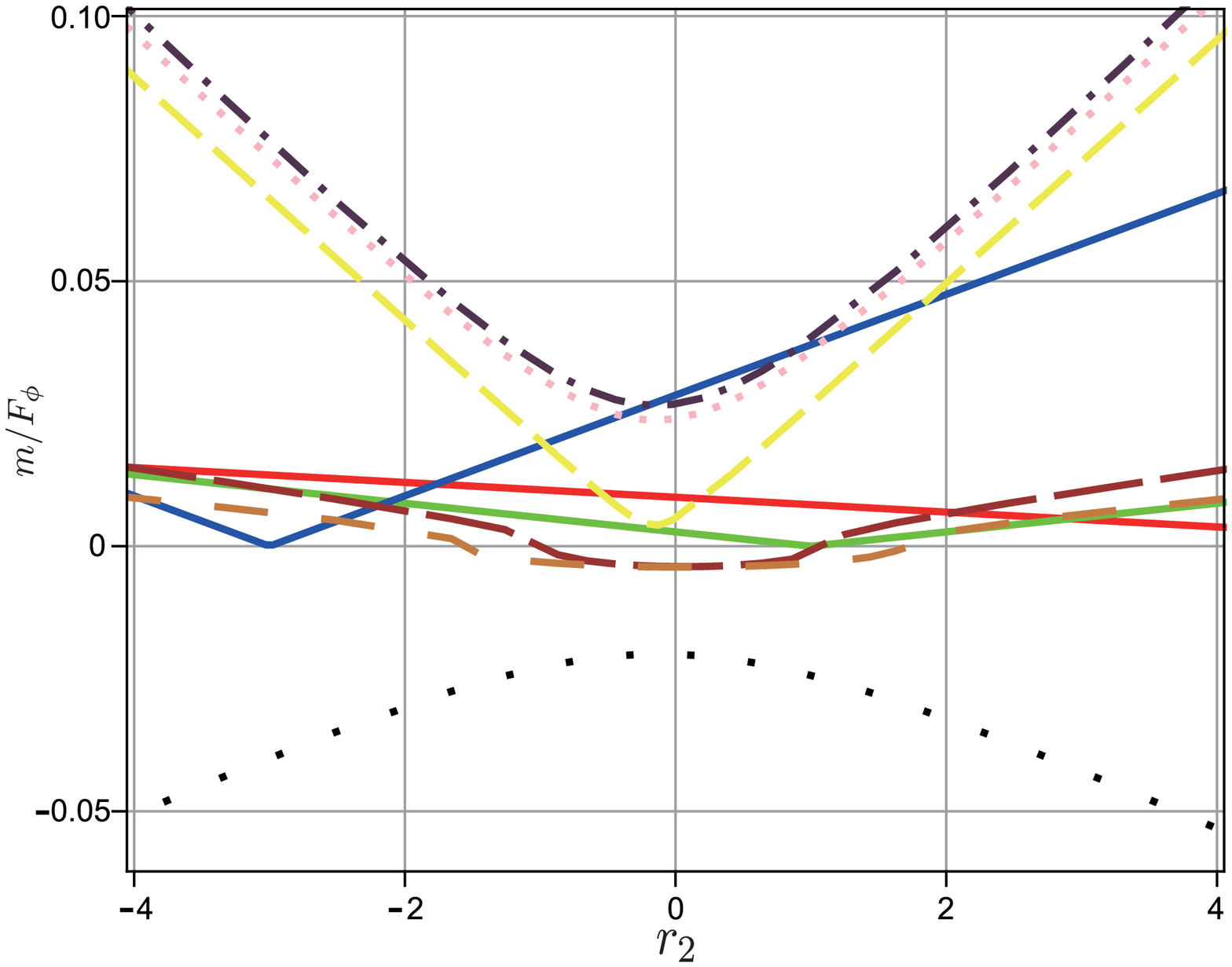}
   \put(-22,26){}
  \end{overpic}
\caption{Same as Figure 1, but for $r_2= 1/2r_3$.}
 \end{center}
\end{figure}

\begin{figure}[htbp]
 \begin{center}
  \begin{overpic}[width=12cm,clip]{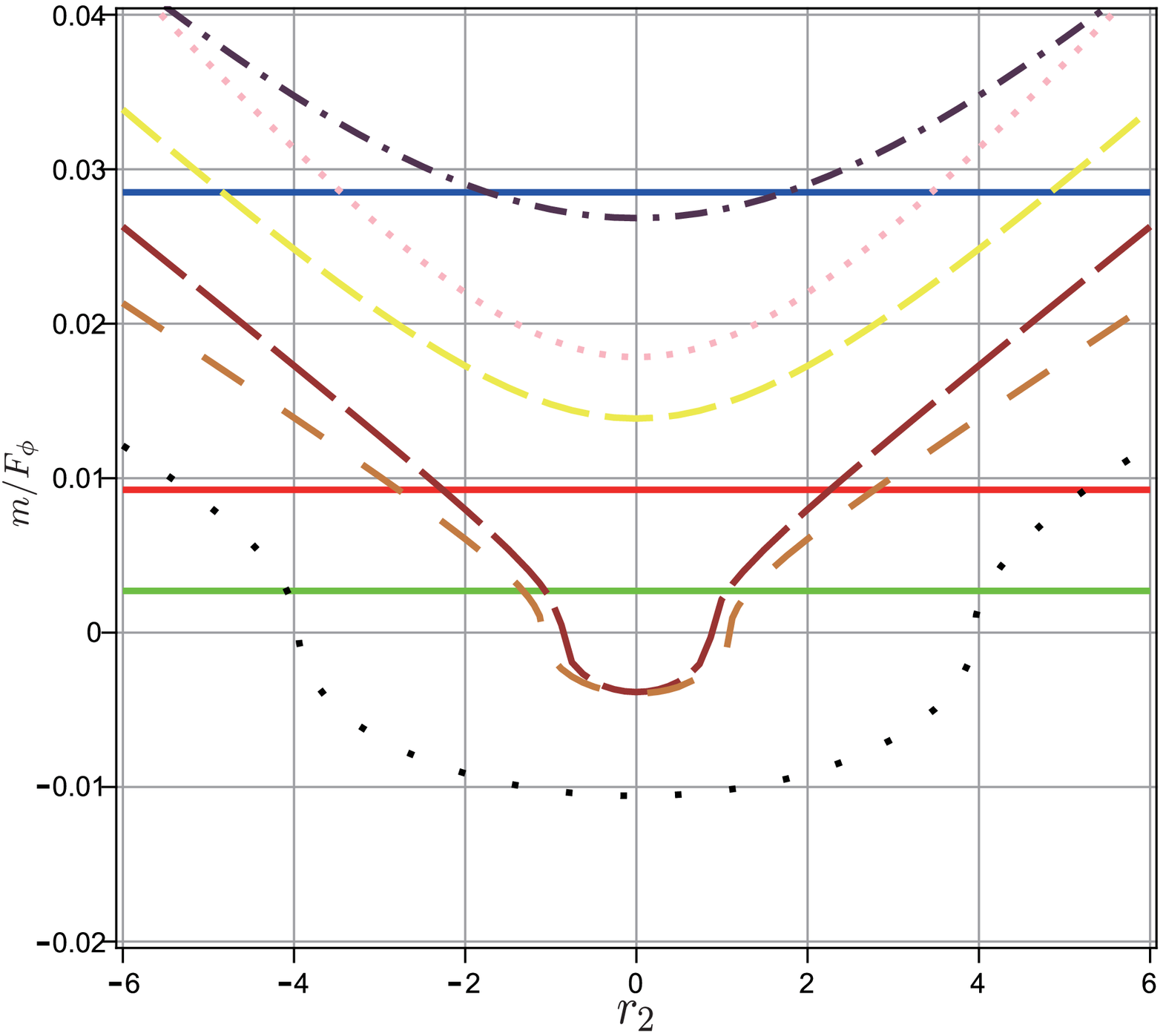}
   \put(-22,26){}
  \end{overpic}
  \begin{overpic}[width=12cm,clip]{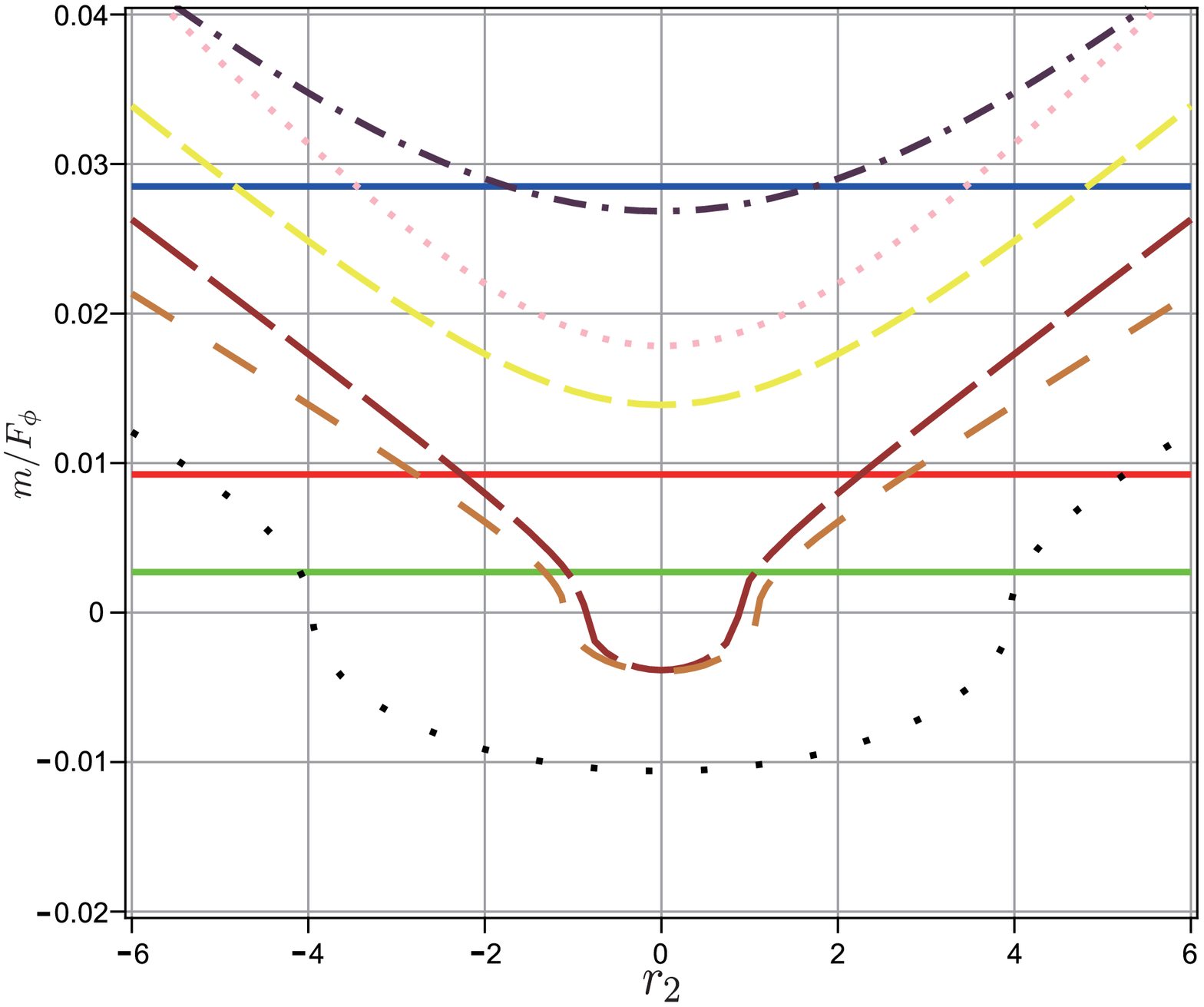}
   \put(-22,26){}
  \end{overpic}
\caption{Same as Figure 1, but we take the messenger scale as $10^{10}$ GeV at the upper panel and $2 \times 10^{16}$ GeV at the lower panel.}
 \end{center}
\end{figure}

\end{document}